# Gravitational Tunneling Radiation

Mario Rabinowitz
Armor Research, lrainbow@stanford.edu
715 Lakemead Way, Redwood City, CA 94062-3922

**Abstract**

The isolated black hole radiation of both Hawking and Zel'dovich are idealized abstractions as there is always another body to distort the potential. This is considered with respect to both gravitational tunneling, and black hole "no-hair" theorems. The effects of a second body are to lower the gravitational barrier of a black hole and to give the barrier a finite rather than infinite width so that a particle can escape by tunneling (as in field emission) or over the top of the lowered barrier (as in Schottky emission). Thus radiation may be emitted from black holes in a process differing from that of Hawking radiation, $P_{SH}$ , which has been undetected for over 24 years. The radiated power from a black hole derived here is $P_R \propto e^{-2\Delta\gamma} P_{SH}$, where $e^{-2\Delta\gamma}$ is the transmission probability for radiation through the barrier. This is similar to electric field emission of electrons from a metal in that the emission can in principle be modulated and beamed. The temperature and entropy of black holes are reexamined. Miniscule black holes herein may help explain the missing mass of the universe, accelerated expansion of the universe, and anomalous rotation of spiral galaxies. A gravitational interference effect for black hole radiation similar to the Aharonov-Bohm effect is also examined.

**Keywords:** Hawking-Zel'dovich Radiation, black holes, gravitational tunneling, universe expansion, galaxy rotation, Aharonov-Bohm effect, hairy black holes, entropy.

Mario Rabinowitz
Armor Research
715 Lakemead Way
Redwood City, CA 94062-3922
e-mail: 715mario@worldnet.att.net
Phone & FAX 650,368-4466

**1. INTRODUCTION**

In 1971 Zel'dovich(1) proposed the first model of radiation from a black hole, "The rotating body [black hole] produces spontaneous pair production [and] in the case when the body can absorb one of the particles, ... the other (anti)particle goes off to infinity and carries away energy and angular momentum." This is quite similar to the model proposed by Hawking(2, 3) in 1974 for radiation from non-rotating black holes. Stephens(4) suggested that the Hawking effect may be more general than just for the gravitational field, and observable in a wide variety of settings. (So that his conclusion does not apply to just a special case, the inverted harmonic potential of Stephens $V(x) = -\frac{1}{2}m\omega^2 x^2$ should have its peak at some value of potential less than zero, rather than at zero as he has it.) However, Hawking radiation has not been observed after over two decades of searching. (5) Scientific papers (6,7, 8) have been written offering reasons why it may not be observable. Belinski(7), an expert in the field, unequivocally concludes "the effect [Hawking radiation] does not exist."

The radiation to be derived herein originates from within a black hole and tunnels out due to the field of a second body (in contrast to Hawking's single body approach). This is similar to electric field emission of electrons from a metal by the application of an external field, except that a replenishing source of mass is not needed since there is only one sign of mass. As in Zel'dovich's and Hawking's approach, quantum mechanics is used to facilitate the analysis which does not relate to a theory of quantum gravity.

As we shall see, tunneling radiation derived here can be emitted at much higher intensities and temperatures in the present epoch than Hawking radiation, and thus may be able to fit the detected universal gamma-ray background(5); and shed light on why Hawking radiation has not yet been observed. The tunneling probability or transmission in this paper, is different than that of Hawking. Hawking et al(9, 10) claim that the pair-creation model of Hawking radiation is equivalent to considering the positive and negative energy particles "as being the same particle which tunnels out from the black hole...." As shown herein, the two processes are not found to be equivalent.

**2. FOUR MODELS OF GRAVITATIONAL QUANTUM TUNNELING**

We could carry through a general abstract solution $e^{-2\Delta\gamma}$ in what follows. For an isolated Einsteinian black hole, depending on angular momentum, there can be a barrier peaked at about 1.5 $R_H$ which is negligible in the zero angular momentum case, where the Schwarzchild or black hole horizon radius, $R_H = 2GM/c^2$. Since the difference between general relativity and the Newtonian gravitational potentials gets small at r greater than 10 $R_H$, for all scales, and as the angular momentum approaches zero, we can calculate specific transmission probabilities using the Newtonian potential in this limit. (Newton's law may fail at very small distances, as does Coulomb's law, due to quantum effects.)

**2.1 Isolated spherical black hole**

The tunneling probability from the gravitational well of an isolated body is zero because the barrier has infinite width. Nevertheless, let us derive it because the solution can give us an insight for the analysis of tunneling in the case of a gravitational potential due to more than one body, where the probability is greater than zero. Tunneling is done non-relativistically to simplify the analysis.

The one-dimensional Schrödinger equation for a mass m in a well of potential energy V due to a spherical body of mass M centered at the origin is

$$\frac{-\hbar^2}{2m}\frac{d^2\Psi}{dr^2} + \Psi[V - E] = \frac{-\hbar^2}{2m}\frac{d^2\Psi}{dr^2} + \Psi\left[\frac{-GmM}{r} - E\right] = 0. \qquad (1)$$

As shown in Fig. 1, the gravitational potential energy of a single isolated spherical body is -GmM/r down to its surface, with total energy $E = -GmM/b_1$ at the classical turning point $b_1$. In general relativity and Newtonian gravity, $R_H$ is the same, and the vacuum gravitational field outside any spherically symmetric object is the same as that of a point mass. A wave function $\Psi$ of the form $\Psi = Ae^{-\gamma(r)}$ is a solution of eq. (1), when $d^2\gamma/dr^2 \approx 0$ is negligible ($\gamma$ is dimensionless).

The tunneling probability between points $b_1$ and $b_2$ is the ratio of probability densities at $b_1$ and $b_2$:

$$\Pi = \frac{\Psi(b_2)\Psi^*(b_2)}{\Psi(b_1)\Psi^*(b_1)} \approx e^{-2[\gamma(b_2)-\gamma(b_1)]} \equiv e^{-2\Delta\gamma} \tag{2}$$

The solution for $\Delta\gamma \equiv \int_{b_1}^{b_2}\left[\frac{2m}{\hbar^2}(V-E)\right]^{1/2} dr$ that satisfies eq. (1) is

$$\Delta\gamma \approx \frac{m}{\hbar}\sqrt{2GM}\left\{\sqrt{\frac{b_2(b_2-b_1)}{b_1}} - \sqrt{b_1}\,\ell n\left[\frac{\sqrt{b_2}+\sqrt{b_2-b_1}}{\sqrt{b_1}}\right]\right\}. \tag{3}$$

Thus as expected $\Pi = 0$, since as $b_2$ approaches $\infty$, $\Delta\gamma$ approaches $\infty$.

**2.2 Elemental Black Hole Facing Another Body**

As derived in Sec. 2.1, the tunneling probability is zero for escape from inside a single isolated black hole, or any isolated body. Let us see what effect an external second body has on the tunneling probability out of a black hole of mass M. Let us take a square well as the simplest approximation to a black hole. The tunneling probability out of an isolated infinite square well is zero; and it may be harder to tunnel out of it than an isolated black hole, as an infinite inward-directed force is encountered at its surface. As shown in Fig. 2, we have a body of mass $M_2$ centered at $R_2$ opposite a black hole centered at the origin of radius $R_H$; and strength of the well $V_0 R_H \approx GM^2$, where $V_0$ is the potential energy depth of the well.

Outside the black hole, we need to solve the Schrödinger equation

$$\frac{-\hbar^2}{2m}\frac{d^2\Psi}{dr^2} + \Psi\left[\frac{-GmM_2}{R_2 - r} - E\right] = 0 \qquad R_H \le r \le \beta_2. \tag{4}$$

The first classical turning point is $R_H$ with $E = -GmM_2/(R_2 - R_H)$, and $E = -GmM/(R_2 - \beta_2)$ at the second classical turning point $\beta_2$.

We solve for $\Delta\gamma$ as before:

$$\Delta\gamma \approx \frac{m}{\hbar}\sqrt{2GM_2}\left\{\sqrt{\frac{(\beta_2 - R_H)(R_2 - R_H)}{R_2 - \beta_2}} - \sqrt{R_2-\beta_2}\,\ell n\left[\frac{\sqrt{R_2-R_H}+\sqrt{\beta_2-R_H}}{\sqrt{R_2-\beta_2}}\right]\right\} \tag{5}$$

As $R_2$ approaches $R_H$ or $\beta_2$ approaches $R_H$ eq. (5) gives $\Delta\gamma$ approaches 0, yielding $\Pi$ approaches 1. This will be the case for all the barriers considered in this paper. (This is not the case in general, as quantum mechanically $\Pi$ may be less than 1 at the top of a well or barrier.) As $R_2$ approaches $\infty$, $\Delta\gamma$ approaches $\infty$, yielding $\Pi$ approaches 0.

Eq. (5) is implicitly equivalent to eq. (3) for a finite $b_2$. This can be seen explicitly by setting $b_1 = R_2 - \beta_2$ and $b_2 = R_2 - R_H$ . It's an example of a general result from the symmetry of quantum tunneling that for a non-absorbing barrier, the transmission amplitude and phase are the same in both directions.(11) This has significance for black hole tunneling in that the transmission probability must be the same into or out of a black hole.

The second body lowers the barrier and gives the barrier a finite rather than infinite width so that a particle can escape by tunneling or over the top of the lowered barrier. Black hole emission is greatest when the companion is a nearby almost black hole, and least when it is a distant ordinary body. The escaping particle may be trapped in the well of the second body. If it is not also a black hole, then escape from it can occur by ordinary processes such as scattering, and gravity-assisted energy from the second body's angular momentum. The second body also helps to validate the tunneling calculation in terms of limits equivalent to an isolated body which yield eq. (3 ).

The use of a square well as a simplified analog of a black hole is intended as a heuristic aid for approximate calculation. It is similar in spirit to the prevalent approach of using a square well to represent the nuclear well where it is impossible to describe by a potential the forces acting on a particle inside the nucleus. A square well model also serves well to heuristically obtain fundamental masses(12); and a cut-off mass for inter-universe tunneling.(13)

**2.3 Black Hole Opposite Another Body**

Now let us look at a less simplified model of a black hole facing a body of mass $M_2$. As shown in Fig. 3, $M_2$ is centered at R opposite a black hole of mass M centered at the origin. Since tunneling is greatest near the top of the barrier, the deviation from a 1/r potential toward the center of each body is not critical. The potentials used are that of two point masses, so $M_2$ may also be a black hole. Thus two little black holes (LBH) may get quite close for maximum tunneling radiation. In this limit, there is a similarity between the analysis here, and what is expected from the Hawking model in that the tidal forces of two LBH add to give more radiation at their interface also producing a repulsive force.

Outside the black hole, we need to solve the Schrödinger equation

$$\frac{-\hbar^2}{2m}\frac{d^2\Psi}{dr^2}+\Psi\left[\frac{-GmM}{r}+\frac{-GmM_2}{R-r}-E\right]=0 \qquad (6)$$

in the region $b_1 \le r \le b_2$, where $b_1$ and $b_2$ are the classical turning points, and $E = -GmM/b_1 + -GmM_2/(R - b_1) = -GmM/b_2 + -GmM_2/(R - b_2)$.

We solve for $\Delta\gamma$ as before:

$$\Delta\gamma \approx \frac{m}{\hbar}\sqrt{\frac{2GM}{d}}\left\{\sqrt{b_2(b_2-d)}-\sqrt{b_1(b_1-d)}-d\,\ell n\left[\frac{\sqrt{b_2}+\sqrt{b_2-d}}{\sqrt{b_1}+\sqrt{b_1-d}}\right]\right\} \qquad (7)$$

where $d= Mb_1(R- b_1)R/[M(R- b_1)R+M_2(b_1)^2]$, and the solution applies for $R >> b_2$, which is valid for $M_2 >> M$.

Eq. (7) reduces to eq. (3) as R approaches infinity as it should, and $\Pi$ approaches zero. One must observe the approximation R much much greater than $b_2$, that was made in deriving eq. (7), when the limit R approaches infinity is taken while holding $R- b_2$ constant, i.e. $b_2$ should also approach infinity when this is done. As in the previous cases $\Delta\gamma$ approaches zero as $b_1$ approaches $b_2$, yielding $\Pi$ approaches 1. When M approaches zero, or $M_2$ approaches infinity, or equivalently $[M/M_2]$ approaches zero, $\Delta\gamma$ approaches zero and $\Pi$ approaches one. Eq. (5) can serve as a lower limit check on $\Pi$ when exact calculations can't be done for the general two-body case. The mass $M_2$ can be fixed in space, orbit around the black hole, or be at R temporarily. Radiation will escape from the black hole as long as $M_2$ 's lingering time is much much greater than the greater of the tunneling time or black hole transit time.

It is remarkable that a black hole of infinite mass in the presence of another body becomes completely transparent quantum mechanically ( $\Pi$ = 1). As an isolated body, a black hole would be completely opaque ($\Pi = 0$). Nevertheless as we shall see, it cannot radiate even when it is completely transparent, since the radiated power is proportional to $[1/M]^2$ which approaches zero as M approaches infinity. The tunneling calculations in Sec. 2 are general and also apply to gravitational tunneling of ordinary bodies.

**2.4 Aharonov-Bohm-like effect (Black hole surrounded by a spherical shell):**

Now let us surround the black hole of Sec. 2.3, with a thin spherical shell of mass $M_2$ at radius $R_2$ as shown in the lower left part of Fig. 4. The potential energy of this system is

$$V = \begin{Bmatrix} \dfrac{-GmM}{r} + \dfrac{-GmM_2}{R_2}, \quad 0 < r \le R_2 \\ \dfrac{-GmM}{r} + \dfrac{-GmM_2}{r} = \dfrac{-Gm(M+M_2)}{r}, \quad r \ge R_2 \end{Bmatrix}. \quad (8)$$

In this case, the Schrödinger equation is

$$\frac{-\hbar^2}{2m}\frac{d^2\Psi}{dr^2} + \Psi[\{V\} - \{E\}] = \frac{-\hbar^2}{2m}\frac{d^2\Psi}{dr^2} + \Psi\begin{bmatrix} \left\{\dfrac{-GmM}{r} + \dfrac{-GmM_2}{R_2}\right\} - \\ \left\{\dfrac{-GmM}{b_1} + \dfrac{-GmM_2}{R_2}\right\} \end{bmatrix} \quad (9)$$

$$= \frac{-\hbar^2}{2m}\frac{d^2\Psi}{dr^2} + \Psi\left[\frac{-GmM}{r} + \frac{GmM}{b_1}\right] = 0$$

which is the same as eq. (1), i.e. the same as if the concentric spherical shell of mass $M_2$ weren't there.

Thus as in Sec. 2.1, the tunneling probability $\Pi = 0$. This is the expected classical result, as the spherical region of constant potential $-GmM_2 / R_2$ exerts no force inside $M_2$ and has no physical consequence by itself. As shown, this is also the case quantum mechanically. Howeve the presence of a third mass such as $M_3$ outside $R_2$ as in Fig. 4, would make the gravitational barrier finite and release radiation; and produce interference at Region B if $M_3$ were also a black hole.

Two systems, each with black hole (B.H.) and concentric shell, opposite each other as in Fig. 4 can not only result in a tunneling probability $\Pi$ greater than 0, but if the potentials were time varying due to M(t), and/or $R_2$(t) and/or $M_2$(t), this can lead to interference effects. M(t), due to the decreasing black hole mass, leads to a time-varying Hamiltonian. Even though $M_2$ produces no gradient in potential, we are not dealing with regions of space with no gradient in the potential (as in the traditional Aharonov-Bohm[14] effect) as there is a gradient due to the black hole inside. However, the time variation of the potential can lead to black hole radiation interference effects at Interference Region A similar to the Aharonov-Bohm effect.

If the time-varying Hamiltonian can be written as the simple sum of the stationary and time-varying part, $H = H_o + V(t)$, then whether or not V is constant in space, $\Psi = \Psi_o e^{-i\phi/\hbar}$ is a solution of the time-dependent Schrödinger equation, $H\Psi = i\hbar\frac{\partial \Psi}{\partial t}$ since

$$i\hbar\frac{\partial}{\partial t}\left[\Psi_o e^{-i\phi/\hbar}\right] = e^{-i\phi/\hbar}\left[i\hbar\frac{\partial \Psi_o}{\partial t} + \Psi_o\frac{\partial \phi}{\partial t}\right] = \left[H_o + V(t)\right]\Psi_o e^{-i\phi/\hbar}, \qquad (10)$$

where $V(t) = \frac{\partial \phi}{\partial t}$. Thus $\phi = \int V(t)dt$, where $\frac{\phi}{\hbar}$ is the phase shift. Without reflectors, Interference Region A is along a line of symmetry between the two black holes. With reflectors, the interference region may be moved to other locations.

**3. TRANSMISSION VS TUNNELING PROBABILITIES**

A distinction must be made between the concepts of transmission and tunneling as used in the terms "transmission probability or transmission coefficient " and "tunneling probability or penetration coefficient." Tunneling probability is a ratio of probability densities and transmission probability is a ratio of probability current densities. For an Einsteinian black hole there is a relatively small barrier at 1.5 $R_H$ that even virtual particles have to tunnel through to get out. Hawking[2,3] uses the terms "tunnel" or "penetrate" through this potential barrier interchangeably, and does not use the term "transmission." Since his calculations or final results for these quantities are not presented, it appears from context that he is using both these terms just for "tunneling probability." The much earlier literature also often did not distinguish between the two concepts. The transmission probability or coefficient

$$\Gamma \equiv \frac{\Psi(b_2)\Psi^*(b_2)v_2}{\Psi(b_1)\Psi^*(b_1)v_1} \approx \frac{v_2}{v_1} e^{-2[\gamma(b_2)-\gamma(b_1)]} = \Pi \frac{v_2}{v_1} \quad , \tag{11}$$

where tunneling is from region 1 (left of the barrier) to region 2 (right of the barrier). $\Gamma = \Pi$ when the velocities $v_1$ and $v_2$ are the same on both sides of the barrier.

In general , without needing an explicit solution for $\Psi$ it can be shown that

$$\Gamma = \frac{\Psi_2 \Psi_2^*}{\Psi_{inc} \Psi_{inc}^*}\left(\frac{v_2}{v_1}\right) = \left[e^{\Delta\gamma} - \tfrac{1}{4} e^{-\Delta\gamma}\right]^{-2} \quad . \tag{12}$$

From eq. (11),

$$\Delta\gamma \equiv \int_{b_1}^{b_2} \left[\frac{2m}{\hbar^2}(V - E)\right]^{1/2} dr. \tag{13}$$

Thus when $\Delta\gamma$ is large, $e^{\Delta\gamma} >> (1/4)e^{-\Delta\gamma}$ in eq. (12), yielding $\Gamma \approx \Pi = e^{-2\Delta\gamma}$. $\Gamma \approx \Pi$ is true in most cases when $b_2 - b_1 >> 0$, and/or $V >> E$. Note that $e^{-2\Delta\gamma}$ is the solution obtained for the various cases in Sec. 2, where $\Delta\gamma$ was obtained via the integral of eq. (13).

However, we shall be mainly interested in the high energy case, when V - E is small which (for our gravitational barriers) implies that the distance between the classical turning points, $b_2 - b_1$ may not be relatively large. At first sight it would appear that we cannot make the approximation $\Gamma \approx \Pi$. Propitiously, the barrier of Sec. 2.3 becomes symmetrical for all energies and barrier widths when $M = M_2$ and then $v_1 = v_2$. Similarly $v_1 \sim v_2$ for $M \sim M_2$. So in this paper $\Gamma \approx \Pi$ is a valid approximation when $M \sim M_2$, and is exactly true for all M and $M_2$ in the case of ultrarelativistic particles, where $v_1 \approx v_2 \approx c$, the speed of light . However, for non-zero rest mass particles, when their energies are low in a non-symmetrical gravitational barrier, this may not be a valid approximation. This seems to have been overlooked by Hawking and others. It is a good approximation for little black holes because low energy particles are a miniscule fraction of the radiation due to extremely high temperatures, but needs to be taken into consideration for intermediate and high mass black holes.

**4. EMISSION RATE**

Since it is generally accepted that black hole radiation is independent of the time-history of the black hole formation, we may for theoretical purposes consider a black hole that has existed for an infinite length of time. Our conclusions about the radiation of such a static everlasting black hole should also be valid for all black holes, despite the virtual inconsistency that any black hole would have evaporated away after an infinite time. Nevertheless, it is legitimate for *gedanken* purposes.

The procedure taken here is similar to that traditionally used for tunneling out of a nucleus. Each approach of the trapped particle to the barrier has the calculated probability of escaping or tunneling through the barrier . Thus we need only know the frequency of approach to the barrier. The time between successive impacts on the barrier for ultrarelativistic particles is

$$\tau = \frac{2\langle r \rangle}{c} \approx \frac{2R_H}{c} = \frac{2\left(2GM / c^2\right)}{c} , \tag{14}$$

where the Schwarzchild radius has been put in for $R_H$. In general relativity time and space exchange roles inside a black hole, so that a particle inexorably falls into the center. Quantum mechanics should allow barrier collisions roughly as given by eq. (14). Asymptotic freedom permits treating the constituents inside an LBH as a hot dilute gas at high energies and short distances. The second body reduces $R_H$ of the LBH primarily in the direction between them.

While there are more circuitous routes exceeding $2R_H$, $\tau$ is likely a good representative collision time for ultrarelativistic particles with the barrier. Thus in the high energy case, the emission rate or probability of emission per unit time from the black hole is

$$\frac{\Gamma}{\tau} = \frac{\Pi}{\tau}\frac{v_2}{v_1} = \frac{\Pi}{\tau}\frac{c}{c} = \frac{\Pi}{\tau} \approx \frac{\Pi c^3}{4GM} = \frac{e^{-2\Delta\gamma}c^3}{4GM} , \tag{15}$$

where $\Delta\gamma$ is given by eqs. (3), (5), and (7) for the models discussed in Section 2. Eq. (15) is a good approximation when the black hole is opposite a large body. A fractional solid angle, $\Delta\Theta / 4\pi$, correction factor needs to multiply eq. (15) when the adjacent body is small.

**5. BLACK HOLE TEMPERATURE**

Let us follow two heuristic approaches in obtaining the temperature, T, of the particles emitted from a black hole. The emitted particles do not undergo a gravitational red shift in tunneling through the barrier. Thus the temperature and average energy of the emitted particles is the same on either side of the barrier, $\langle E \rangle = \langle E_e \rangle \approx kT$. The simplest approach relates the momentum, p, of an ultrarelativistic particle inside a black hole, in this case a square well of width $2R_H$, to its de Broglie wavelength, $\lambda$, where $\lambda / 2 \approx 2R_H$. Hence:

$$p = \frac{h}{\lambda} \approx \frac{h}{4R_H} = \frac{h}{4\left(\frac{2GM}{c^2}\right)} = \frac{hc^2}{8GM}, \text{ and} \tag{16}$$

$$T \approx \frac{\langle E_e \rangle}{k} = \frac{E}{k} = \frac{pc}{k} = \left[\frac{hc^3}{8kG}\right]\frac{1}{M}. \tag{17}$$

In the second simple dynamical approach, we can use semi-classical Bohr theory to obtain T for a hydrogenic-like bound particle of mass m to a black hole of mass M for a 1/r gravitational potential as in Fig. 1.[15] The energy of the Bohr orbit is:

$$E = \frac{2\pi^2 (GMm)^2 m}{n^2 h^2}. \tag{18}$$

The radius of the Bohr orbit is

$$r = \frac{n^2 \hbar^2}{GMm^2}. \tag{19}$$

At the top of the well, $r = R_H = 2GM / c^2$, yielding

$$n^2 = \frac{2G^2 M^2 m^2}{\hbar^2 c^2}. \tag{20}$$

Substituting eq. (20) into (18)

$$E = \frac{2\pi^2 (GMm)^2 m}{h^2}\left[\frac{\hbar^2 c^2}{2G^2 M^2 m^2}\right] = \frac{1}{4} mc^2. \tag{21}$$

(Note that the energy $E \propto mc^2$ was obtained from the escape velocity ≥ c from a black hole without invoking relativity.)

From eq. (21), we have the temperature of the emitted particles

$$T \approx \frac{\langle E_e \rangle}{k} = \frac{E}{k} = \frac{mc^2}{4k} = \left(\frac{h\upsilon}{c^2}\right)\frac{c^2}{4k} \approx \left(\frac{h}{4k}\right)\frac{c}{\lambda} = \left(\frac{hc}{4k}\right)\frac{1}{4R_H} = \left[\frac{\pi\hbar c^3}{16kG}\right]\frac{1}{M}. \qquad (22)$$

This is ~ T in eq. (17), indicating that T is roughly model independent as obtained here.

The Hawking 1974 value for temperature[1] is a factor of 2 smaller than his 1975 value. This is not critical, and the 1975 expression[(2)] is

$$T = \left[\frac{\hbar c^3}{4\pi kG}\right]\frac{1}{M} = \left[2.46x10^{26}\right]\left(\frac{1}{M}\right){}^{o}K \quad , \qquad (23)$$

with M in gm. For M ~ $10^{15}$ gm (the largest mass that can survive to the present for Hawking), T ~ $10^{11}$ K. As we shall see, my theory permits the survival of much smaller masses such as for example M ~ $10^9$ gm with T ~ $10^{17}$ K.

Both eqs. (17) and (22) are close to the Hawking expression for temperature derived on the basis of entropy considerations with an important exception. No correction for gravitational red shift needs to made here since the particles tunnel through the barrier without change in energy. The Hawking temperature appears to have an inconsistency. Although originally proposed as not being real,[(16)] this temperature is now asserted and generally accepted as being the gravitationally red shifted temperature. As shown by eq. (24), their new view implies an infinite temperature at the horizon of all black holes, since the red shift goes to zero as measured at large distances from any hole if the surface temperature were finite. For the real temperature, they averred "the effective temperature of a black hole is zero ... because the time dilation factor [red shift] tends to zero on the horizon." Temperature could be inferred for an LBH from the energy distribution of emitted particles. Red shift would greatly reduce the frequencies of the observed gamma rays. It is not clear that this has been taken into consideration.

Particles that originate at or outside the horizon of an isolated black hole must lose energy in escaping the gravitational potential of the black hole. For example, in the case of photons

$$\nu_{\text{observer}} = \nu(r_{\text{emission site}})\left[1-\frac{2GM/c^2}{r}\right]^{1/2} = \nu(r_{\text{emiss site}})\left[1-\frac{R_H}{r}\right]^{1/2}, \quad (24)$$

where $\nu_{\text{observer}}$ is the frequency detected by the observer at a very large distance from the black hole, and $\nu(r_{\text{emission site}})$ is the frequency at the radial distance r from the center of the black hole where the particle pair was created outside the black hole. Eq. (24) does not depend on the gravitational potential between the emission site and observer, but only on the emission proximity to $R_H$. We see from eq. (24) that for $r = R_H$ and any finite $\nu(r_{\text{emission site}})$, $\nu_{\text{observer}} = 0$ implying that any finite temperature at $R_H$ must red shift to zero as measured at large distances from a black hole.

Nevertheless, the 1975 Hawking temperature(3) will be used here because it is widely accepted; because the dynamical temperature inside the black hole derived here is close to the Hawking temperature; and because by having the same temperature as a starting point we can clearly see how our other results differ. Temperature is not critical here, as it is to Hawking, as the radiation does not have to be black body. So $\langle E_e \rangle$ could be used here instead of T.

## 6. BLACK HOLE TUNNELING POWER

The power radiated by tunneling from a black hole of volume $\Omega$ is

$$P_R = \frac{\int \Psi^* E_e \frac{\Gamma}{\tau} \Psi d\Omega}{\int \Psi^* \Psi d\Omega} \approx \left\langle E_e \frac{\Pi}{\tau} \right\rangle \sim \langle E_e \rangle \frac{\langle e^{-2\Delta\gamma} \rangle c^3}{4GM}, \quad (25)$$

where $\langle E_e \rangle \approx kT$ is the average energy of the emitted particle. So combining eq. (25) with eq. (23) [or (22) or (17)] for the tunneling radiation power:

$$P_R \approx \left[\frac{\hbar c^3}{4\pi GM}\right] \frac{\langle e^{-2\Delta\gamma} \rangle c^3}{4GM} = \left[\frac{\hbar c^6 \langle e^{-2\Delta\gamma} \rangle}{16\pi G^2}\right] \frac{1}{M^2}. \quad (26)$$

Note that $P_R$ was obtained without invoking field fluctuations, pair creation, etc.

The Hawking radiation power, $P_{SH}$, follows the Stefan-Boltzmann radiation power density law $\sigma T^4$, when $\frac{8\pi GME}{\hbar c^3} >> 1$. For Hawking :

$$P_{SH} \approx 4\pi R_H^2\left[\sigma T^4\right] = 4\pi\left(\frac{2GM}{c^2}\right)^2 \sigma\left[\frac{\hbar c^3}{4\pi kGM}\right]^4 = \frac{\hbar^4 c^8}{16\pi^3 k^4 G^2}\{\sigma\}\left[\frac{1}{M^2}\right]. \quad (27)$$

where $\sigma$ is the Stefan-Boltzmann constant. Although $P_R$ and $P_{SH}$ appear quite disparate, the differences almost disappear if we substitute into eq. (27) the value obtained for $\sigma$ by integrating the Planck distribution over all frequencies:

$$\sigma = \left\{\frac{\pi^2 k^4}{60\hbar^3 c^2}\right\}, \quad (28)$$

$$P_{SH} = \frac{\hbar^4 c^8}{16\pi^3 k^4 G^2}\left\{\frac{\pi^2 k^4}{60\hbar^3 c^2}\right\}\left[\frac{1}{M^2}\right] = \frac{\hbar c^6}{16\pi G^2}\left\{\frac{1}{60}\right\}\left[\frac{1}{M^2}\right]. \text{ Thus} \quad (29)$$

$$P_R = 60\left\langle e^{-2\Delta\gamma}\right\rangle P_{SH}\,. \quad (30)$$

Note that even though $P_R \propto T$ and $P_{SH} \propto T^4$, they can be put into an equivalent form, aside from the numerical factor $60\left\langle e^{-2\Delta\gamma}\right\rangle$. The form of eq. (26) suggests that $P_{SH}$ may be interpreted as the radiation inside the black hole, and $P_R$ is the part that tunnels out.

Hawking's (1974) value[2] for T is a factor of 2 smaller, and since it enters into $P_{SH}$ as $2^{-4}$, we would have gotten

$$P_{R2} = 960\left\langle e^{-2\Delta\gamma}\right\rangle P_{SH}. \quad (31)$$

The text refers only to "temperature is of the order of ", and the context of these papers does not differentiate between the 1974 and 1975 values.

**7. BLACK HOLE LIFETIME AND SURVIVAL MASS**

The evaporation rate for a black hole of mass M is $d\left(Mc^2\right) / dt = -P_R$, which gives the lifetime

$$t = \frac{16\pi G^2}{3\hbar c^4\left\langle e^{-2\Delta\gamma}\right\rangle}\left[M^3\right]. \quad (32)$$

This implies that the smallest mass that can survive up to a time t is

$$M_{small} = \left( \frac{3\hbar c^4 \left\langle e^{-2\Delta\gamma} \right\rangle}{16\pi G^2} \right)^{1/3} \left[ t^{1/3} \right]. \tag{33}$$

Primordial black holes with M >> $M_{small}$ have not lost an appreciable fraction of their mass up to the present. Those with M << $M_{small}$ would have evaporated away long ago.

Thus the smallest mass that can survive within ~ $10^{17}$ sec (age of our universe) is

$$M_{small} \geq 10^{15} \left\langle e^{-2\Delta\gamma} \right\rangle^{1/3} \text{ gm} \tag{34}$$

Inasmuch as $0 \leq e^{-2\Delta\gamma} \leq 1$, an entire range of black hole masses much smaller than $10^{15}$ gm may have survived from the beginning of the universe to the present than permitted by Hawking's theory. For example, if the average tunneling probability $\left\langle e^{-2\Delta\gamma} \right\rangle \sim 10^{-9}$, then $M_{small} \sim 10^{12}$ gm and these bodies will presently radiate ~ $\left[10^{15} / 10^{12}\right]^2 = 10^6$ times more power than a $10^{15}$ gm black hole with the same $e^{-2\Delta\gamma}$.

The above-mentioned differences in the expected radiation may help in understanding why the Hawking radiation profile(5) doesn't fit the observed gamma-ray background. The present theory may help in understanding the gamma-ray background which has far more photons at higher photon energy than expected from Hawking's model.

## 8. ACCELERATED EXPANSION OF THE UNIVERSE

Little black holes are excellent candidates for the missing mass of the universe, which is up to 95% of the $10^{56}$ gm mass of the universe. (13) Although they are quite hot, they qualify as cold dark matter since their velocities << c. One piece of evidence that we do not know what 95% of the universe is made of, comes from spiral galaxies as first observed in 1983 by the unheralded Vera Rubin.(17) The missing mass gives the stars ~ constant linear velocities around their galactic center independent of radial distance r, rather than the expected Keplerian velocities $\propto 1/\sqrt{r}$. Other evidence comes from clusters of galaxies. The speeds at which individual galaxies are moving in these clusters are so high that the clusters would fly apart unless they were held together by a stronger gravitational attraction than provided by the masses of all the galaxies. Further evidence comes from the universal microwave background radiation. The radiation from Hawking little black holes would likely have either interfered with early universe nucleosynthesis or broken up products of nucleosynthesis after the nuclear reactions were over. Furthermore, that many of Hawking little black holes would fry the universe.

The little black hole radiation model derived herein is much less likely to interfere with nucleosynthesis than Hawking's because the radiation is much reduced, and is beamed rather than omnidirectional. Thus these quiescent holes can be much smaller than previously considered since the power radiated is proportional to the transmission coefficient and inversely proportional to the mass squared. Thus these primordial black holes would be much less subject to the limits imposed by nucleosynthesis arguments than the Hawking model. The baryonic matter in them would have bypassed the deuterium and helium formation that occurred during the era of nucleosynthesis. That stars orbit with ~ constant linear velocity may result from the increase in total mass of little black holes with radial distance from a galactic center due to radiation reaction force driving them outward by a similar analysis to that starting with eq. (35), as well as a lower evaporation rate at larger radial distances.

This mechanism may also be able to account for the recently observed accelerated expansion of the universe, in the analysis shown below. One of the most remarkable and exciting discoveries of 1998 revealed that the universe is accelerating in its expansion.[18, 19] The implication is that the universe is older, bigger, and less dense than previously thought. This discovery was unanticipated, and has had no explanation. It calls long-standing cosmological theories into question. It may shed light on the enigma that some stars appear to be older than the previously accepted age of the universe.

It had been thought that the expansion was either at a constant rate, or decelerating due to gravity. If little black holes represent a substantial fraction (up to 95%, because in terms of “smoothness” LBH can be mistaken for energy, since they are so small they can be “smoothed” over”) of the missing mass of the universe, then their radially directed inward radiation in interacting with the universe as the second body is a good candidate model for the accelerated expansion of the universe. It appears that only Einstein's cosmological constant and inflation have been considered as explanations thus far. However a big cosmological constant makes the vacuum enormously more massive than is consistent with observation or quantum theory. Directed radiation from little black holes is a possible explanation that does not have these problems.

Let us determine the maximum acceleration observable from the earth due to the radiation reaction force experienced by a little black hole of mass M in a spherical shell at radius $R_U$, surrounding mass $M_U$ of the universe. For a first approximation, let us ignore the deceleration caused by gravity. Bodies at the edge of the universe are moving out radially near the speed of light, thus requiring relativistic treatment. As seen from the earth, the acceleration is

$$a \equiv \frac{dv}{dt} = \frac{-c}{M}(1-\beta^2)\frac{dM}{dt}, \tag{35}$$

where $\beta \equiv v / c$, and $\frac{dM}{dt} = \frac{-P_R}{c^2}$. $P_R$ is given by eq. (26) ; $\Delta\gamma \approx \frac{m}{\hbar}\sqrt{\frac{2GM}{d}}\{b_2 - b_1\}$ from eq. (7) since $b_2$ and $b_1$ are both >> d; and $d \approx MR_U / M_U$ as given just below eq. (7) since $R_U >> b_2$ and $M_U >> M$. Thus eq.(35) becomes

$$a = \frac{(1-\beta^2)}{M^3}\left[\frac{\hbar c^5}{16\pi G^2}\right]\left\langle \exp-\left[c\frac{b_2 - b_1}{2\pi M}\sqrt{\frac{2M_U}{GR_U}}\right]\right\rangle. \tag{36}$$

Taking $da/dM = 0$, in eq.(36) we can find the relationship between M and $(b_2 - b_1)$ for maximum acceleration,

$$\frac{(b_2 - b_1)}{M} = \left[\frac{c}{6\pi}\sqrt{\frac{2M_U}{GR_U}}\right]^{-1}. \tag{37}$$

Substituting eq. (37) in eq. (36) greatly simplifies the exponential, giving the maximum possible acceleration:

$$a_{max} \le \frac{(1-\beta^2)}{M^3}\left[\frac{\hbar c^5}{16\pi G^2}\right]e^{-3} \tag{38}$$

If the total mass of a spherical shell of the universe is dominated by an ensemble of little black holes, then their acceleration will transport the rest of the bodies in the shell with them by gravitational attraction. The acceleration of each shell will not exceed the value given by eq. (38). Interestingly, even though $R_U$ was used to derive eq. (38), it is independent of $R_U$ . Correction for the deceleration caused by gravity depends on radial distance.

Strictly speaking, a spherically symmetric universe is only rigorously homogeneous with respect to its center. However, if it is infinite , this would not show up in experimental measurements such as the uniformity in all directions of the microwave background radiation. If the universe is finite, measurements made at distances relatively close to its center (compared with its radial size) would tend to hardly show the inhomogeneity. In this case, the inhomogeneity of off-center observations would tend to be lost in the experimental uncertainties. LBH would also yield accelerated expansion for a non-Euclidean universe since each LBH would be repelled from its neighbors by the beamed radiation. Expansion in all geometries gives the appearance than any point may be considered to be central as all points appear to be expanding away from all others.

**9. ENTROPY AND BLACK HOLES**

Bekenstein[20] found that the entropy of a black hole is

$$S_{bh} = kAc^3 / 4G\hbar = k\left[\frac{M}{M_{Pl}}\right]^2 = k \ln N, \tag{39}$$

where A is its surface area/$4\pi$ (neglecting the warpage of space near LBH) M is the mass of the black hole, $M_{Pl} = 2.18 \times 10^{-5}$ gm is the Planck mass, and the right hand side is the standard Boltzmann statistical mechanical entropy of a system containing N distinct states. It follows from his formulation that the entropy of black holes is tremendously greater than the entropy of ordinary bodies of the same mass. For example, our sun of mass $2 \times 10^{33}$ gm has entropy $S \approx 10^{42}\,\mathrm{erg/K}$, whereas a black hole of the same mass has entropy $S_{bh} \approx 10^{60}\,\mathrm{erg/K}$, $10^{18}$ times higher. If the universe were 95% full of such black holes, there would be $10^{23}$ of them with an excess entropy of $10^{41}$ times that of our universe filled with stars like our sun. Thus there is a colossally higher probability that the big bang produced black holes dominantly over ordinary matter. This is a possible solution to the conundrum of why the early universe appears to have so little entropy. It appears likely that a large percentage of the mass of at least the primordial universe was composed of little black holes according to my model. This is particularly so, since interference with nucleosynthesis would no longer be an issue. Another less compelling, but simpler argument can also be made: The time just subsequent to the big bang is a time of extremely high densities of mass-energy which is just the state of little black holes. If the equation of state of primeval high density stuff is not too hard, one may well expect LBH to be a major constituent of the remnants of the big bang. For Hawking the Universe can only be filled with about one-millionth of its mass with LBH, or there would be too much Hawking radiation.

Some insights may be gained by rewriting the area in eq. (39) in terms of the Schwarzchild radius,

$$S_{bh} = \frac{kc^3}{4G\hbar} R_H^2 = \frac{kc^3}{4G\hbar}\left[\frac{2GM}{c^2}\right]^2 = \frac{kGM^2}{\hbar c}. \tag{40}$$

Substituting in eq. (40) for the black hole mass M in terms of the 1974 temperature as given by Hawking[(2)]

$$S_{bh} = \frac{kGM^2}{\hbar c} = \frac{kG}{\hbar c}\left[\frac{\hbar c^3}{8\pi kGT}\right]^2 = \frac{\hbar c^5}{64\pi^2 Gk}\left[\frac{1}{T^2}\right], \tag{41}$$

with Planck's constant appearing explicitly.

If the tunneling radiated power of my model as given by eq. (26) is substituted into eq. (40),

$$S_{bh} \propto k\frac{c^5}{G}\left[\frac{e^{-2\Delta\gamma(\hbar)}}{P_R}\right], \tag{42}$$

where Planck's constant also appears in the transmission coefficient $e^{-2\Delta\gamma(\hbar)}$, as one would expect for all quantum mechanical entities.

Similarly substituting for the black hole mass M in equation (40) in terms of eq. (27) for the Hawking radiated power ,

$$S_{bh} \propto k\frac{c^5}{G}\left[\frac{1}{P_{SH}}\right]. \tag{43}$$

Equation (43) is interesting, as it appears that for a given radiated power $P_{SH}$, the entropy is independent of Planck's constant.

The entropy equations for black holes illustrate an aspect of entropy that seems to be peculiar with both Einsteinian and Newtonian gravity, and many other cases. Gravitational systems tend to exhibit a negative heat capacity. Black holes oddly get hotter (rather than cooler) with a local decrease and global increase in entropy, the more energy they lose by radiation. This is ascribable to quantum mechanics because as the hole shrinks, the quantum wavelength must also decrease leading to a higher momentum and hence higher temperature. A classical black hole should not shrink by evaporation because it supposedly cannot radiate. But if it could, it would get cooler. Dunning-Davies and Lavenda have analyzed the concept of entropy with great clarity and examine other perplexing thermodynamic anomalies associated with the Hawking black hole model.[21]

The following examples show that black holes are not the only things that unexpectedly increase their temperature. A non-quantum mechanical Newtonian example would be satellites losing energy as they move through the atmosphere causing them to gain kinetic energy (at the loss of overall energy and gravitational potential energy) as they fall towards the Earth. This is because they lose angular momentum due to the torque exerted by air resistance. This would also be true in the analogous macroscopic electrostatic case. For a high density ensemble of macro-particles, having a high collision frequency as they fall into a central force as they orbit in a viscous medium, their effective non-equilibrium temperature will increase. In a Joule-Thomson expansion, one usually observes a lowering of the temperature because the total energy is conserved, and in most cases the potential energy increases with the expansion leading to a lower kinetic energy and hence lower temperature. However, at high temperature and/or high pressure, expansion can decrease the potential energy, and increase the temperature.

The above satellite example does not apply to satellites upon which the dissipative force exerts no torque. Lunar tidal friction decreases both the Earth's and the Moon's energy, but hardly if at all exerts torque on the Moon to decrease its angular momentum. In fact the Moon's angular momentum may even increase as the spin momentum of the Earth decreases. Even if the Moon's net angular momentum were just conserved, tidal dissipation would cause the Moon to move further from the Earth with a decrease in kinetic energy.

## 10. HAIRY BLACK HOLES

Black hole "no-hair" theorems state that time-independent, non-rotating symmetrical black holes can be completely characterized by a few variables such as mass and charge associated with long-range gauge fields.(22) These theorems indicate that in the process of collapse, asymmetries will be radiated away essentially as gravitational radiation.(23) Therefore, the symmetry breaking by the presence of a second body which causes the black hole to become irregular, presents a predicament in the survival lifetime of little black holes. The simple gravitational radiation fix cannot remove the perturbation and restore symmetry without the entire evaporation of the black hole. Similarly, the lowering of the barrier by the second body would appear to permit classical escape from a black hole. These questions need further deliberation.

## 11. DISCUSSION

Einsteinian black holes have an effective potential energy far from the hole proportional to 1/r, the same as Newtonian black holes. The differences between the two are only near the black hole and inside. Hartle and Hawking(9) have calculated the tunneling of radiation out of a black hole by the Feynman path-integral method in support of Hawking's(2, 3) earlier approach. Concerns related to tunneling out of Einsteinian black holes may be avoided by considering the black holes herein to be Newtonian. Einstein himself was troubled with the nature of black holes in General

Relativity. At present there is no direct or indirect experimental evidence concerning the regions very near or inside a black hole. The model derived here allows for indirect testing of the nature of black holes through the observed radiation.

The second body both lowers and thins the barrier allowing radiation to escape (~ 37 % by electrons and positrons, ~ 8 % by photons, and the remainder equally divided by the six kinds of neutrinos together with a very small component of gravitons) from inside a black hole over the lowered barrier, or by tunneling as considered in this paper. Since there is relatively little antimatter in our universe, it would not be surprising if for similar reasons, antimatter tunnels out of black holes in smaller quantities than matter. After the radiated particles have tunneled through the barrier, because these particles have a negative total energy they are trapped in the well of the second body. There are a number of ordinary energy excitation processes by which these particles can escape an ordinary second body such as scattering and gravitationally assisted slingshot escape similar to that given to space probes where the particle gains energy at the expense of the orbital or spin energy of the second body. Momentum transfer excitation is a simple process for escape from the second body. For example if a particle of mass $M_3$ (orbiting the second body) has a head-on collision with the tunneling particle of mass m, the kinetic energy of m can increase by about a factor of 9 if both particles have approximately equal and opposite velocities, and $M_3 >> m$, which may be sufficient to excite m into a positive energy state and escape from the second body. It is straightforward to escape the second ordinary body, but without a lowered barrier only tunneling escape is possible from the black hole. Of course, some particles will tunnel back into the hole since $e^{-2\Delta\gamma}$ is the same at the same energy from the second body into the black hole, but a lower density of particles and energy degradation in the second well greatly reduces the transmission rate back into the hole.

Though black holes were long considered to be a fiction, they now appear to be accepted by most astrophysicists.[(24)] There is an apparent incompatibility between general relativity and quantum mechanics, as Bohr-Sommerfeld quantum mechanics is antithetical to the equivalence principle.[15] Neither Hawking nor this paper deal with quantum gravity. Hawking basically deals with a single isolated black hole. To my knowledge, the general relativistic solution for a black hole in the presence of other bodies has not been derived as yet. I would expect that the black hole radius is not only a function of its own mass, but is also a function (in different directions) of the mass of nearby bodies. The two or more body problem is at the heart of my calculation, as the second body thins and lowers the barrier. Mashhoon[25] analyzes limitations which can be helpful in considering the multi-faceted problems related to black hole radiation.

## 12. CONCLUSION

A goal of this paper has been to present an alternative model which can be experimentally tested since Hawking radiation has proven elusive to detect. This paper gives an insight as to why this may be so, as the radiated power from a black hole derived here $P_R \propto e^{-2\Delta\gamma} P_{SH}$ depends strongly on the factor $e^{-2\Delta\gamma}$. In this epoch, for a given mass, the tunneling probability for emitted particles can result in a radiated power many orders of magnitude smaller than that calculated by Hawking.[2,3] This lower radiated power permits the survival of much smaller black holes from the early black hole dominated universe to the present which have much higher temperatures and hence much higher energy photons.[26, 27] These have the potential of matching the observed gamma-ray background, and helping to understand anomalous galactic rotation. Questions have been raised regarding black hole “no hair” theorems and the temperature of a black hole as measured at the black hole.

Little black holes may be able to account for the missing mass of the universe; and possibly ball lightning.[28-30] In a very young compact universe, radially directed tunneling radiation would have been substantial and may have contributed to the expansion of the early universe. In later epochs this radially directed radiation can help clarify the recently discovered accelerated expansion of the universe.

**Acknowledgment**

I wish to thank Ned Britt, Art Cohn, Mark Davidson, Felipe Garcia, and Fred Mayer for helpful discussions.

Received 17 February 1999.

**Résumé**

La radiation d'un trou noir isolé de Hawking et Zel'dovich sont des théories idéalisées, étant donné qu'il y a toujours un autre corps à défigurer le potentiel. On considère celui-ci en ce qui concerne l'effet tunnel de la gravitation et aussi les théorèmes des trous noirs "sans cheveux." Les effets d'un deuxième corps font abaisser sans symétrie la barrière de la gravitation d'un trou noir et donner à la barrière une largeur finie plutôt qu'infinie afin qu'une particule puisse s'échapper par percer un tunnel ou bien par dessus de la barrière abaissée. Ainsi la radiation peut s'émettre des trous noirs d'une méthode qui diffère de celle de la radiation Hawking, qu'on n'a pas perçue pendant plus de 24 ans. L'énergie émise par la radiation d'un trou noir que l'on retire de la présente est $P_R \propto e^{-2\Delta\gamma} P_{SH}$ où $e^{-2\Delta\gamma}$ est la probabilité de la transmission de la radiation à travers la barrière. Elle est semblable à l'émission des électrons d'un métal sur le champ électrique, de manière que l'on peut en principe moduler et diriger l'émission. On réexamine la température des trous noirs. Les trous noirs minuscules ci-examinés peuvent expliquer la matière manquante de l'univers, l'expansion accélérée de l'univers, et la rotation anormale des galaxies spirales. On examine aussi un effet de l'intervention de la gravitation pour la radiation des trous noirs semblable à l'effet Aharonov-Bohm.

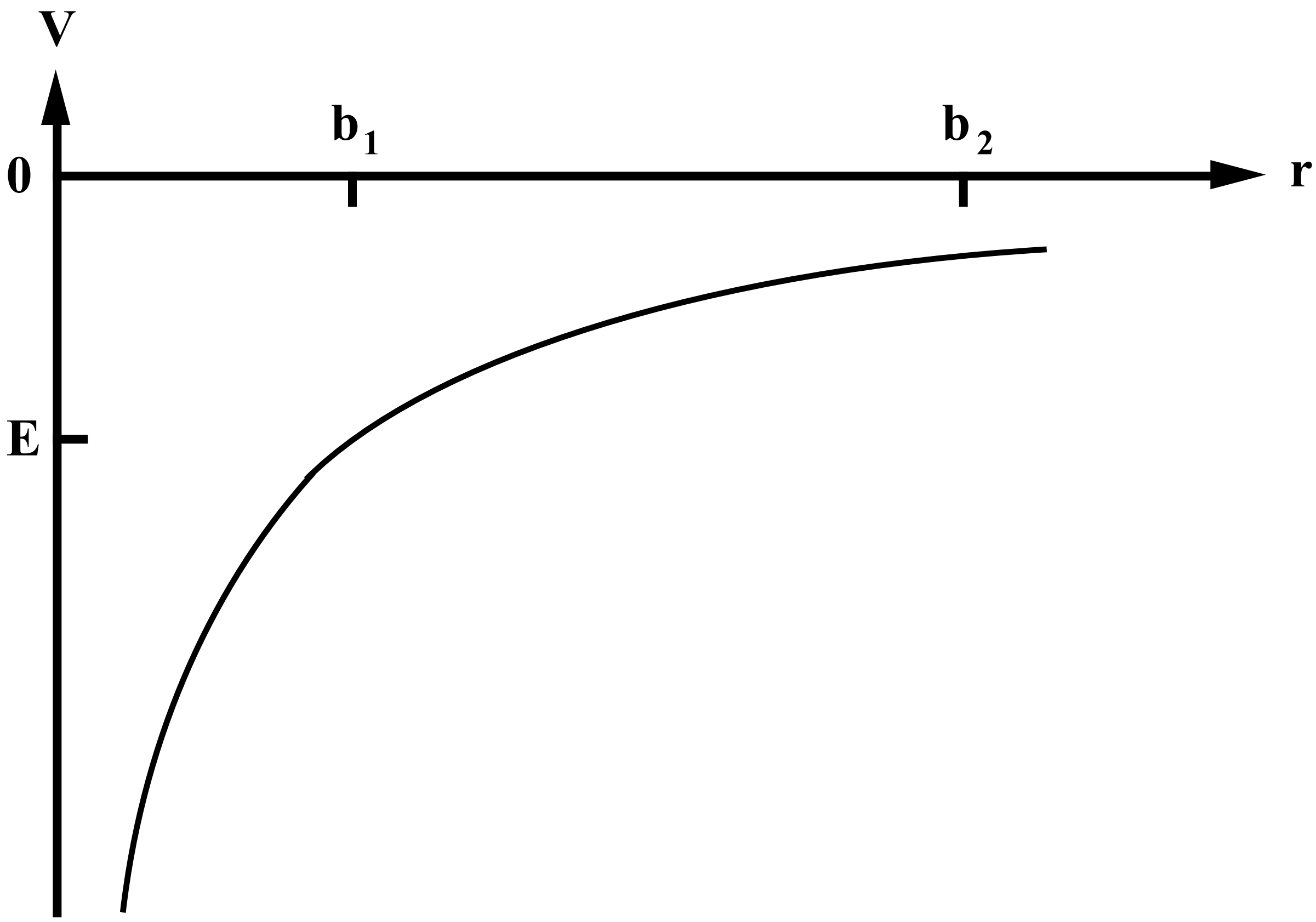

**Fig. 1.** Spherically symmetric gravitational potential energy of an isolated body.

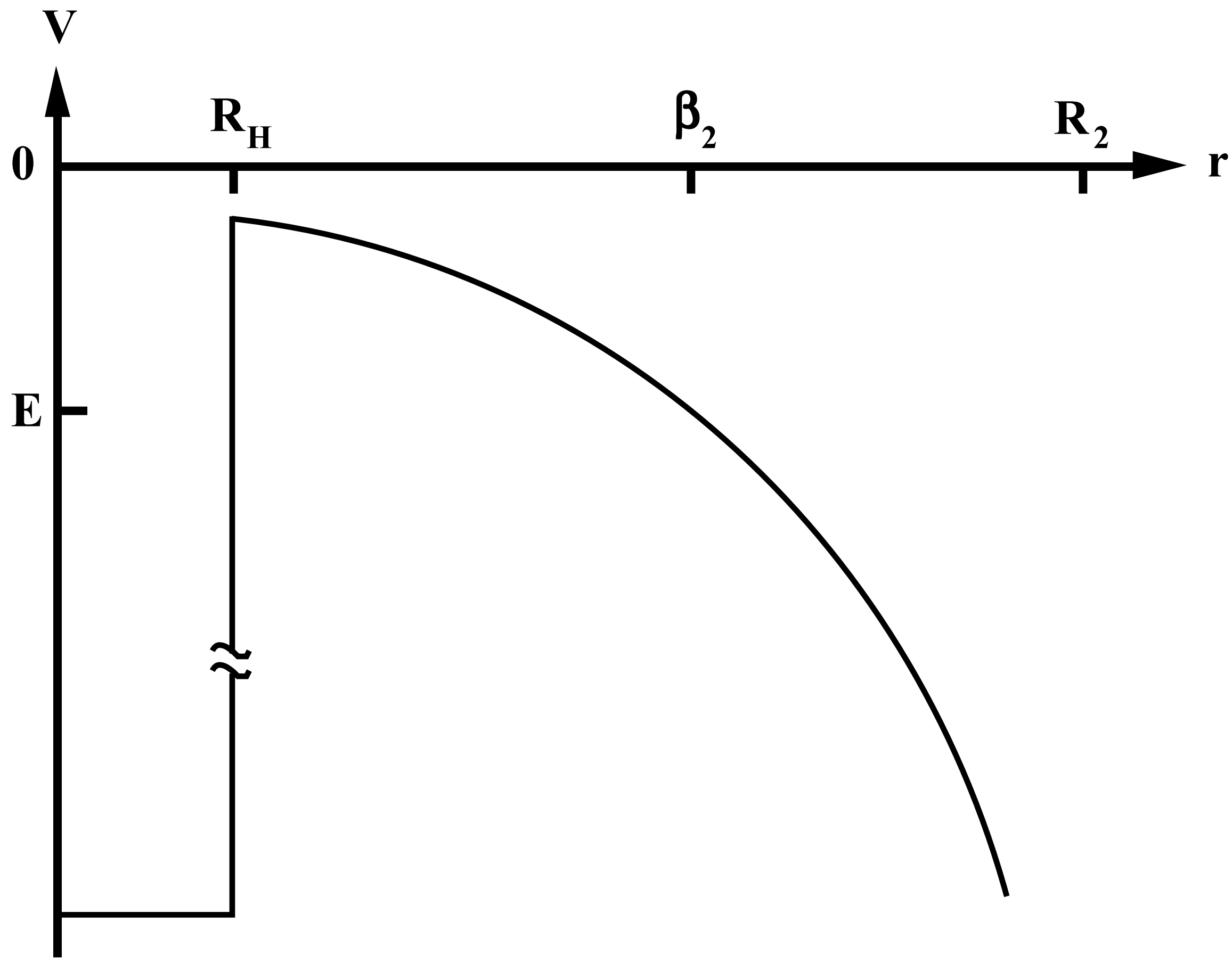

**Fig. 2.** Gravitational potential energy of mass $M_2$ at $R_2$ , opposite a black hole of radius $R_H$ , represented by a square well.

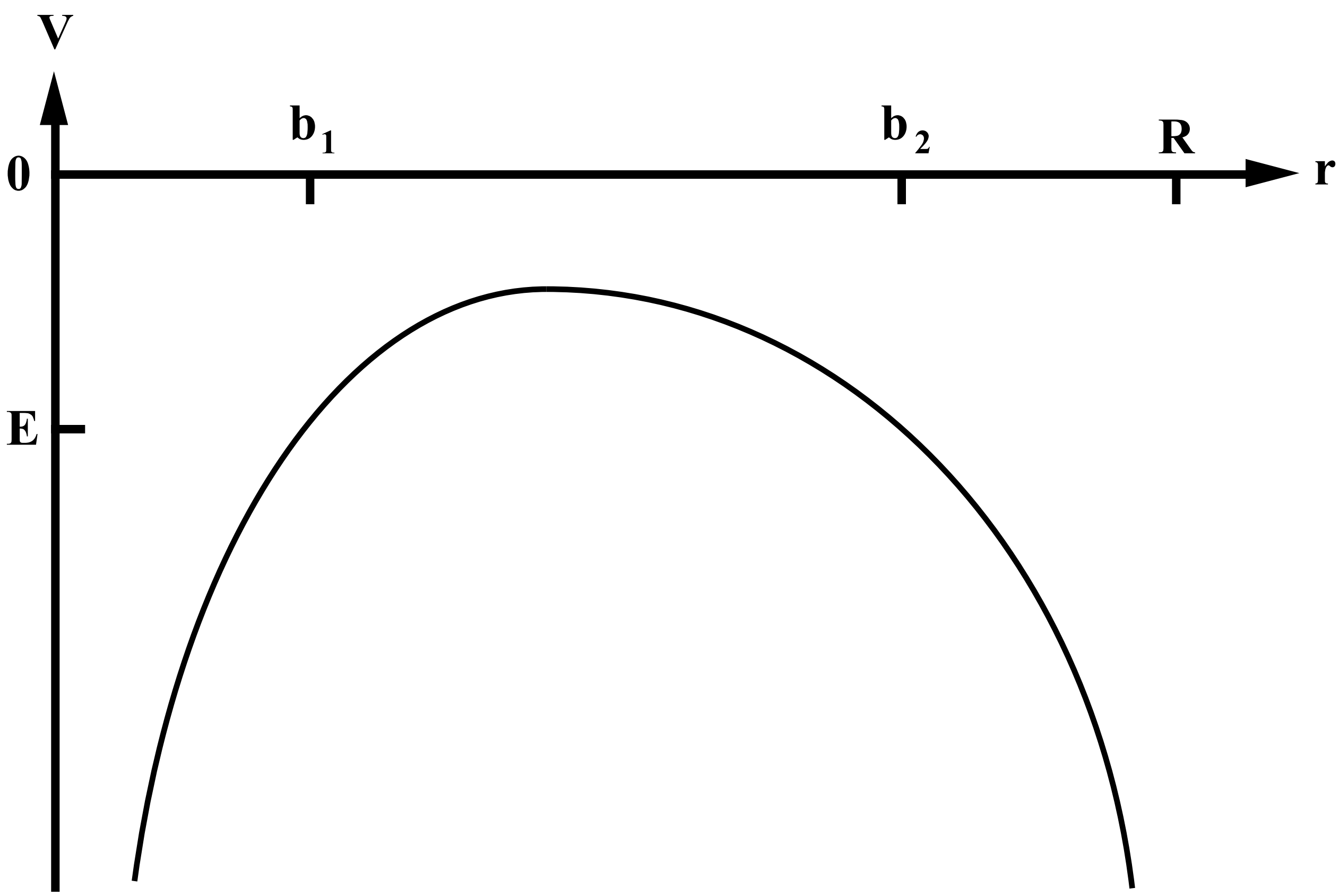

**Fig. 3.** Gravitational barrier resulting from mass $M_2$ at R opposite a black hole of mass M at the origin.

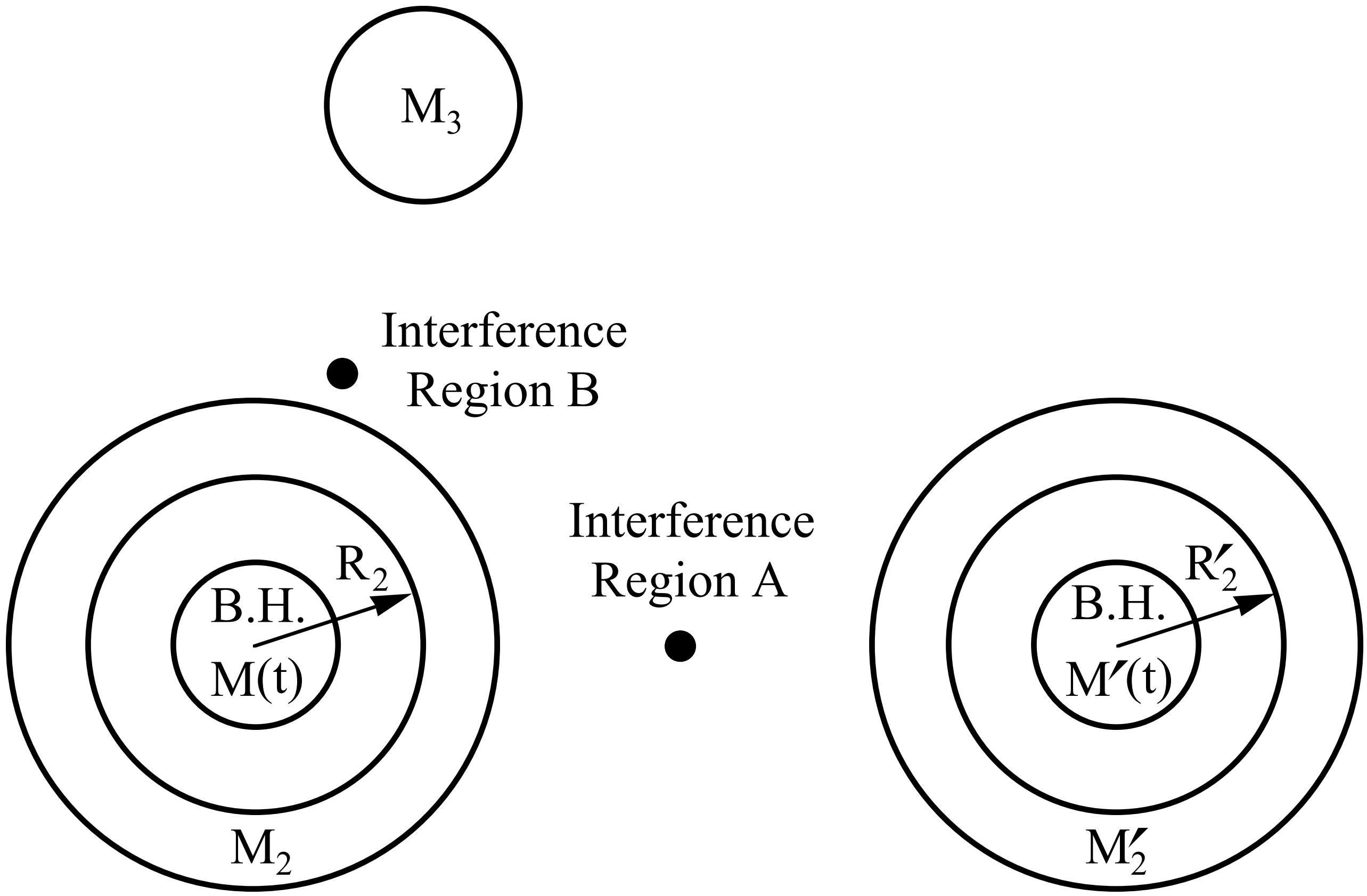

**Fig. 4.** Schematic representation of Aharonov-Bohm-like interference with time-varying gravitational potentials of black holes (B.H.).